\documentclass[12pt]{article}
\usepackage{amsmath}
\usepackage{graphicx,psfrag,epsf}
\usepackage{enumerate}
\usepackage{natbib}
\usepackage{textcomp}
\usepackage[hyphens]{url} 
\usepackage{hyperref}

\newcommand{\blind}{0}

\usepackage{color}
\usepackage{fancyvrb}

\DefineVerbatimEnvironment{Highlighting}{Verbatim}{commandchars=\\\{\}}
\usepackage{framed}
\definecolor{shadecolor}{RGB}{248,248,248}
\newenvironment{Shaded}{\begin{snugshade}}{\end{snugshade}}

\newcommand{\AttributeTok}[1]{\textcolor[rgb]{0.77,0.63,0.00}{#1}}

\newcommand{\CommentTok}[1]{\textcolor[rgb]{0.56,0.35,0.01}{\textit{#1}}}

\newcommand{\DecValTok}[1]{\textcolor[rgb]{0.00,0.00,0.81}{#1}}

\newcommand{\FunctionTok}[1]{\textcolor[rgb]{0.00,0.00,0.00}{#1}}

\newcommand{\NormalTok}[1]{#1}

\newcommand{\OtherTok}[1]{\textcolor[rgb]{0.56,0.35,0.01}{#1}}

\newcommand{\SpecialCharTok}[1]{\textcolor[rgb]{0.00,0.00,0.00}{#1}}

\newcommand{\StringTok}[1]{\textcolor[rgb]{0.31,0.60,0.02}{#1}}

\providecommand{\tightlist}{%
  \setlength{\itemsep}{0pt}\setlength{\parskip}{0pt}}

\usepackage{longtable,booktabs,array}
\usepackage{calc} 
\usepackage{etoolbox}
\makeatletter
\patchcmd\longtable{\par}{\if@noskipsec\mbox{}\fi\par}{}{}
\makeatother
\IfFileExists{footnotehyper.sty}{\usepackage{footnotehyper}}{\usepackage{footnote}}
\makesavenoteenv{longtable}

\usepackage{graphics}
\usepackage{float}

\begin{document}

\def\spacingset#1{\renewcommand{\baselinestretch}%
{#1}\small\normalsize} \spacingset{1}


\if0\blind
{
  \title{\bf Teaching Visual Accessibility in Introductory Data Science Classes with Multi-Modal Data Representations}

  \author{
        JooYoung Seo \\
    School of Information Sciences, University of Illinois Urbana-Champaign\\
     and \\     Mine Dogucu \\
    Department of Statistical Science, University College London\\
Department of Statistics, University of California Irvine\\
      }
  \maketitle
} \fi

\if1\blind
{
  \bigskip
  \bigskip
  \bigskip
  \begin{center}
    {\LARGE\bf Teaching Visual Accessibility in Introductory Data Science Classes with Multi-Modal Data Representations}
  \end{center}
  \medskip
} \fi

\bigskip
\begin{abstract}
Although there are various ways to represent data patterns and models, visualization has been primarily taught in many data science courses for its efficiency. Such vision-dependent output may cause critical barriers against those who are blind and visually impaired and people with learning disabilities. We argue that instructors need to teach multiple data representation methods so that all students can produce data products that are more accessible. In this paper, we argue that accessibility should be taught as early as the introductory course as part of the data science curriculum so that regardless of whether learners major in data science or not, they can have foundational exposure to accessibility. As data science educators who teach accessibility as part of our lower-division courses in two different institutions, we share specific examples that can be utilized by other data science instructors.
\end{abstract}

\noindent%
\vfill

\newpage
\spacingset{1.45} 

\hypertarget{introduction}{%
\section{Introduction}\label{introduction}}

According to LinkedIn's U.S. Emerging Jobs Report, data scientists rank among the top emerging jobs \citeyearpar{linkedin}.
With the urgent need for training a higher number of skilled data scientists, many institutes of higher education are developing their own data science curricula.
However, a consensus on what should be included in such curricula has not yet been reached.
One of the few guidelines on the topic is written by the ACM Data Science Task Force and titled \emph{Computing Competencies for Undergraduate Data Science Curricula}. In this report, accessibility is a ``foundational consideration'' at the knowledge level when it comes to displaying data \citep{danyluk2021computing}.

Training on accessibility is also an important job-readiness aspect of curriculum design. The Partnership on Employment \& Accessible Technology (PEAT) states in their Accessible Technology Skills Gap Report \citeyearpar{peat} that 84\% of industry correspondents that they work with say ``it is important or very important for them to hire developers and designers with accessible technology skills.''
Even though this statistic may be different specifically for data scientists, the data science field is broad with the intersection in careers as developers (e.g., data dashboard and package developers) and designers (e.g., data visualization experts).
The PEAT also states that 60\% of industry respondents ``it was difficult or very difficult for their organization to find job candidates with accessibility skills.''

Despite the recent recommendations, accessibility from disabilities perspectives has not yet found its place in the data science course materials.
One main reason for this could be that accessibility is often covered in the computer sciences \citep{kawas2019teaching} or the digital arts \citep{barata2019inclusion}, if at all covered.
However, data science is an interdisciplinary field that is taught in various disciplines, including computer science, statistics, business schools, political science, and biological sciences \citep{schwab2021data}.
Thus, we believe that the resources and conversations around accessibility have not yet extended to the broader data science education community at the scale that they should.
For instance, many introductory data science courses \citep{yan2019first, baumer2015data}, textbooks \citep{adhikari2019computational, wickham2016r}, or curricula \citep{schwab2021data} either fail to mention accessibility at all or only focus on the accessing data.

In this manuscript, we argue that accessibility should be taught early in the data science curriculum so that regardless of whether learners major in data science or not, they can have foundational exposure to accessibility.
As data science educators who teach accessibility as part of our lower-division courses in two different institutions, we share specific examples that can be utilized by other data science instructors.
We teach R \citep{R-base} as the main language in our courses; thus, our examples are shaped around using R. However, instructors using a different set of language(s) or even teaching language-agnostic courses may still find the content applicable for their own courses.

\hypertarget{learning-goals-and-objectives}{%
\section{Learning Goals and Objectives}\label{learning-goals-and-objectives}}

Although there are various ways to represent data patterns and models, visualization has been primarily taught in many data science courses for its efficiency \citep{kimAccessibleVisualizationDesign2021}.
Such vision-dependent output may cause critical barriers against blind and visually impaired people and sometimes people with learning disabilities \citep{kimAccessibleVisualizationDesign2021, marriottInclusiveDataVisualization2021, leeReachingBroaderAudiences2020}.
We argue that instructors need to teach multiple data representation methods so that all students can produce data products that are more accessible.
To this end, we have considered multi-data representations and their relation to accessibility in our courses, and that's what we will demonstrate throughout this paper.

In incorporating accessibility to our course contents, we also wanted our students to be familiar with assistive technologies, and since our focus has been on visual accessibility, it was a natural fit to include screen readers.
Screen reader is an assistive technology that supports blind or visually impaired people in using their computer by reading aloud the contents. Recently, screen reader has been provided as one of the built-in accessibility features on operating systems, such as Windows Narrator and Apple's VoiceOver.

However, we also want to make our teaching as inclusive as possible to other types of dis/abilities so that we are not just confined to issues of visual disabilities.

Therefore, our curricular goal is to broadly ensure that the current and the next generation of data scientists provide public-facing outputs, including websites and analysis reports, in accessible forms and representations for people with diverse abilities. To achieve this goal, we set three specific and minimal learning objectives like below.

Students should:

\begin{enumerate}
\def\labelenumi{\arabic{enumi}.}
\tightlist
\item
  get familiar with Americans with Disabilities Act;
\item
  use at least one assistive technology;
\item
  consider different representations of data.
\end{enumerate}

The first objective provides our students with a chance to learn legislative concepts and empathy building concerning diversity issues. As both the authors teach data science in the United States, it was a natural choice for us to make students aware of the Americans with Disabilities Act (ADA).
The ADA is a civil rights law that prohibits discrimination against individuals with disabilities.
Regardless of their own disability status, as future data scientists, we expect our students to produce data science products that are accessible to all, not only because it is the right thing to do but also because they may have future responsibilities as data scientists to comply to ADA.

The second learning objective echoes our argument on accessible data representation beyond one-type modality. People with disabilities may rely on different types of assistive technology in their day-to-day life.
Assistive Technology is any form of technology (e.g., software, hardware) that helps people with disabilities perform certain activities. Examples of assistive technology include, but are not limited to, walking sticks, wheelchairs, and screen readers. As we noted above, we introduced screen reader as one of these examples when our students were working on a data visualization project; however, we acknowledge that various assistive technologies, such as speech recognition and video captioning, could be integrated into data science courses in diverse contexts (e.g., data collection and data presentation contexts).

It is important to note that the first two objectives are not data science specific but are related to accessibility on a broader scale.
The third objective, however, is data science specific. In the remaining section of this paper, we will focus further on the third objective by illustrating how we have taught multiple data representations.

\hypertarget{data-visualization}{%
\subsection{Data Visualization}\label{data-visualization}}

Data visualization is the most common representation of data. Despite their use by many, data visualizations are not accessible to all.
Colors are an important part of data visualizations and an aspect that can limit accessibility for color-blind people.
When teaching data visualizations, we include three basic rules for incorporating colors into data visualization. Students should be able to

\begin{enumerate}
\def\labelenumi{\arabic{enumi}.}
\tightlist
\item
  simulate color blindness (if they are sighted and not color blind);
\item
  pick color blind friendly colors;
\item
  not only rely on color for differentiating data.
\end{enumerate}

\begin{figure}
\includegraphics{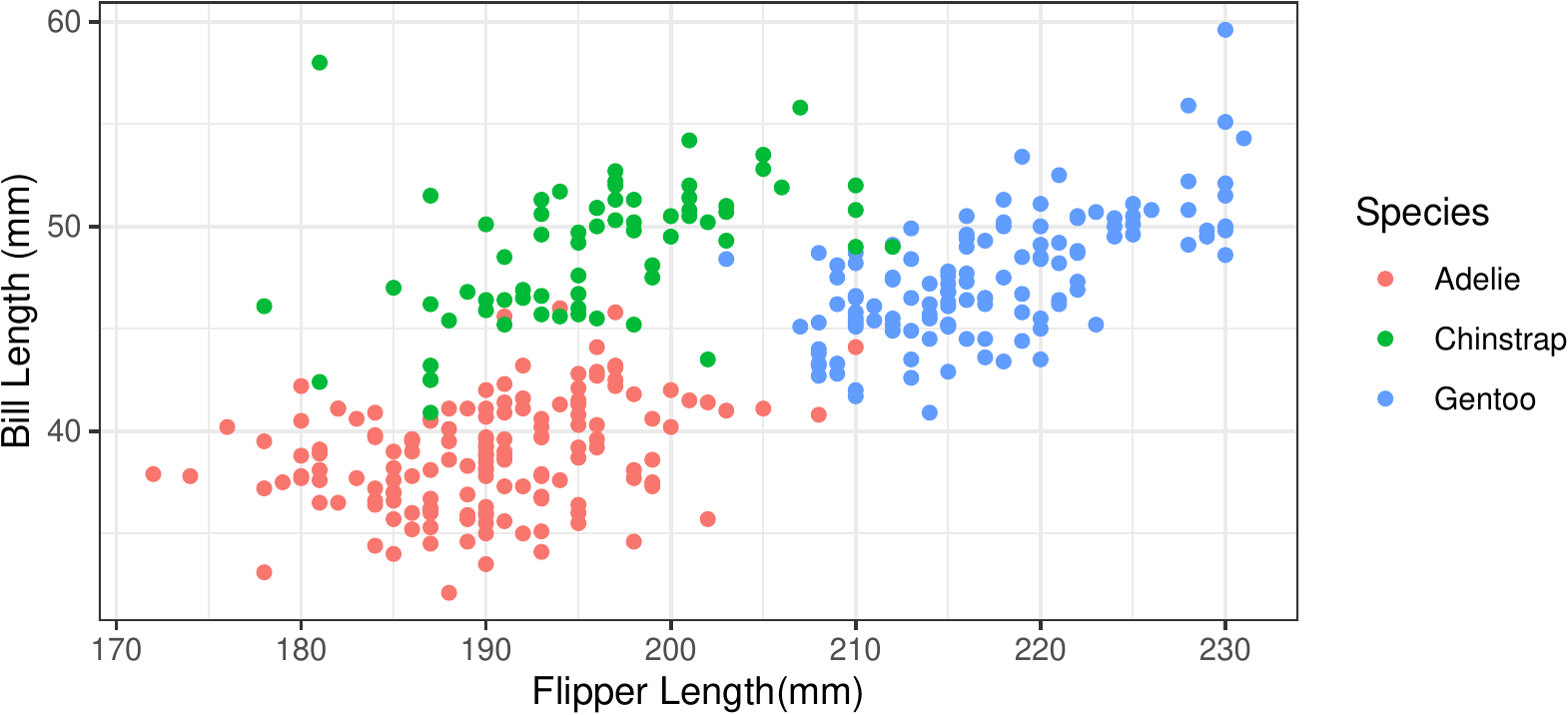} \caption{Sample scatterplot showing the relationship between flipper length (mm) and bill length (mm) of penguins for three different species of penguins}\label{fig:penguins-basic}
\end{figure}

We show an example of incorporating these three basic rules by using the Palmer Penguins data \citep{R-palmerpenguins}.
We visualize the relationship between flipper length and bill length of penguins and color the data points differently for each species in Figure \ref{fig:penguins-basic}.

The \texttt{colorblindr} package \citep{R-colorblindr} provides a set of functions that can be useful in teaching and practicing color blindness accessibility.
For instance, \texttt{colorblindr::cvd\_grid()} function provides color-deficiency simulations of a given plot.
In the code chunk below, the plot is named as \texttt{fig}.
The code creates a grid of four plots where each plot shows a different color vision deficiency, including deuteranomaly, protanomaly, tritanopia, and desaturated.
This grid allows those without any color vision deficiencies to experience what data visualizations would look like to a sighted person that has color defiency.

\begin{Shaded}
\begin{Highlighting}[]
\NormalTok{colorblindr}\SpecialCharTok{::}\FunctionTok{cvd\_grid}\NormalTok{(fig)}
\end{Highlighting}
\end{Shaded}

\begin{figure}
\includegraphics{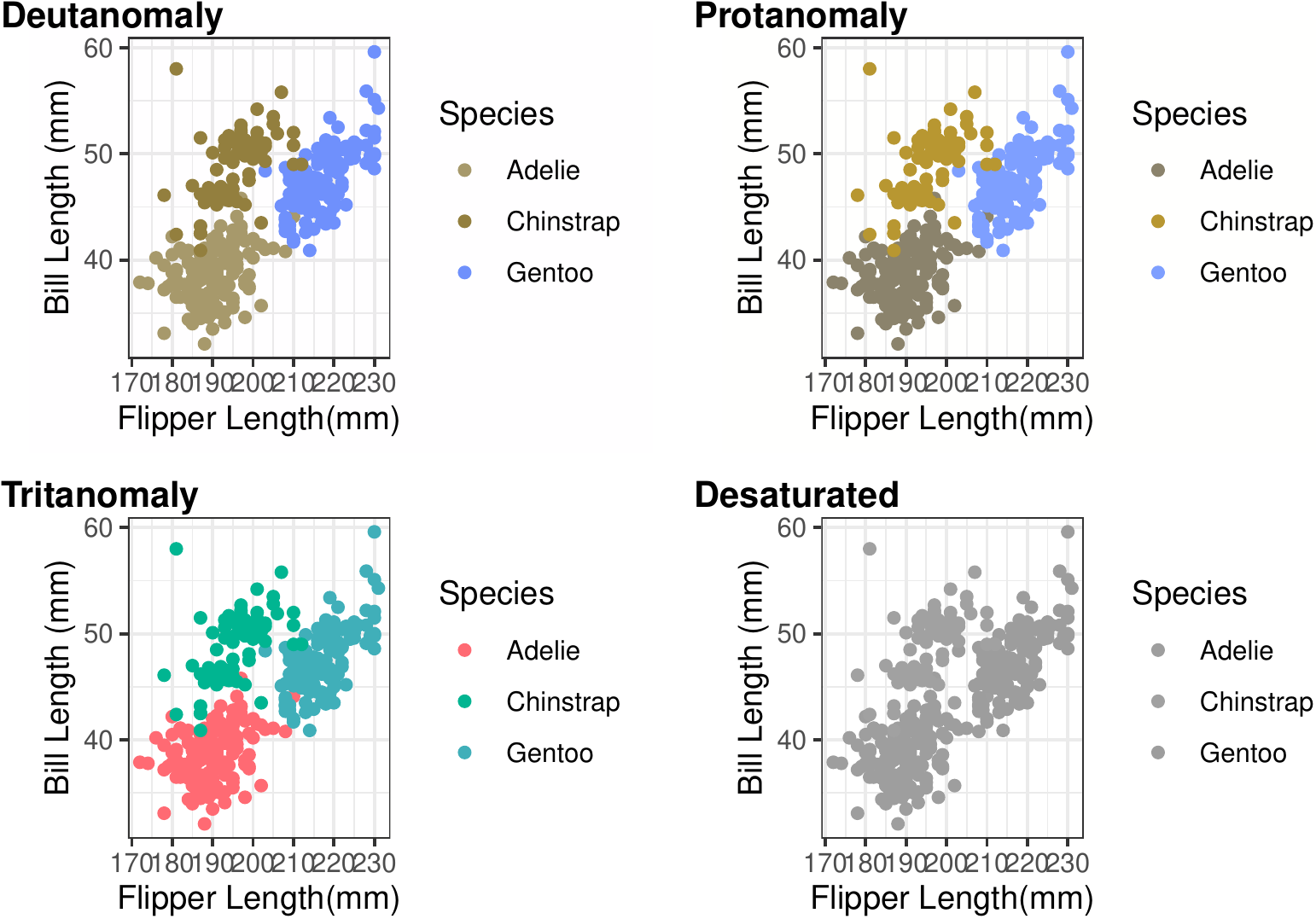} \caption{Color Blindness Simulation with colorblindr}\label{fig:penguins-cvd}
\end{figure}

\begin{figure}
\includegraphics{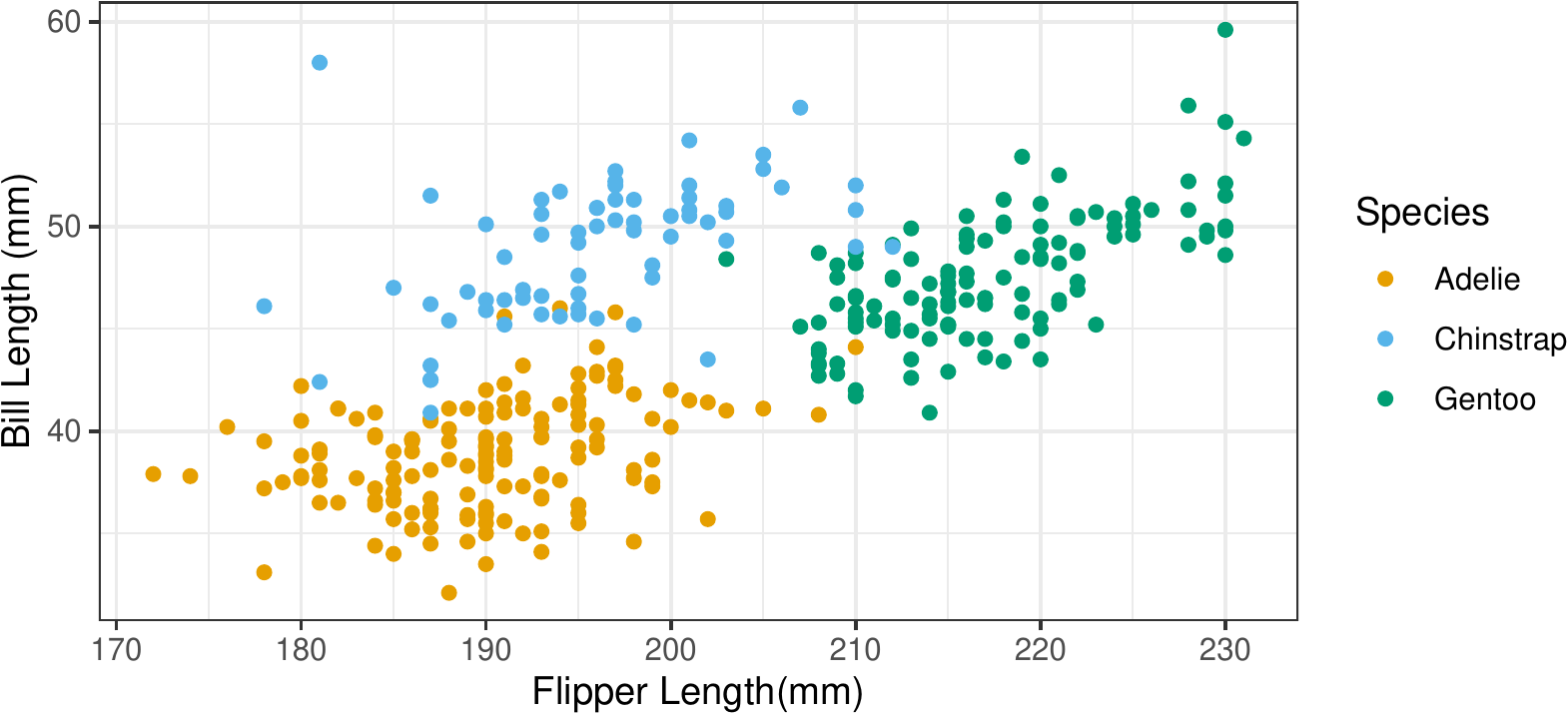} \caption{Scatterplot Using Okabe-Ito Color Pallette}\label{fig:penguins-okabe}
\end{figure}

For their data visualizations, instead of relying on the default color selection of R packages, students are introduced to the Okabe-Ito color palette, which is known to be accessible to people with color vision deficiencies \citep{okabe}.
One can find specific colors in the pallette by the R code \texttt{palette.colors(palette\ =\ "Okabe-Ito")}.
In addition, students can be introduced to \texttt{colorblindr::scale\_color\_OkabeIto()} and \texttt{colorblindr::scale\_fill\_OkabeIto()} to automatically switch to using this color pallette by adding a layer to a ggplot presentation.
For instance, the plot in Figure \ref{fig:penguins-basic} can be switched to the Okabe-Ito color palette as shown in Figure \ref{fig:penguins-okabe} by using the code below.

\begin{Shaded}
\begin{Highlighting}[]
\NormalTok{fig }\SpecialCharTok{+}\NormalTok{ colorblindr}\SpecialCharTok{::}\FunctionTok{scale\_color\_OkabeIto}\NormalTok{()}
\end{Highlighting}
\end{Shaded}

Last but not least, students should learn not to rely only on color to differentiate data points, in this case, Species.
For instance, they can be introduced to shapes and have data points that are circles, triangles, and squares for each species.
This can also be extended to line graphs as having continuous or dashed lines.
Facetting can also be utilized.
For instance, each of the species could have its own scatterplot side-by-side.

In this manuscript and in our introductory courses, we focus on static data visualizations.
Those who teach interactive data visualizations may also consider covering cognitive and motor accessibility.
Our focus here is on visual accessibility, but it is worth noting that all subdomains of accessibility are relevant to data science learners and can be incorporated into the curriculum as seen fit.

\hypertarget{data-verbalization-i.e.-alt-text}{%
\subsection{Data verbalization (i.e., alt text)}\label{data-verbalization-i.e.-alt-text}}

Our definition of ``data verbalization'' refers to a way of representing data patterns using verbal description with alternative text (i.e., alt text).
Images without alt text markup are not readily accessible to assistive technologies (i.e., screen readers; refreshable braille displays; text-based voice commands) and web search parsers.
However, appropriately designed alt text can not only give assistive technology users minimum access to visualized data but also provide all students with a much richer context \citep{lundgardAccessibleVisualizationNatural2022}.

\begin{figure}
\includegraphics{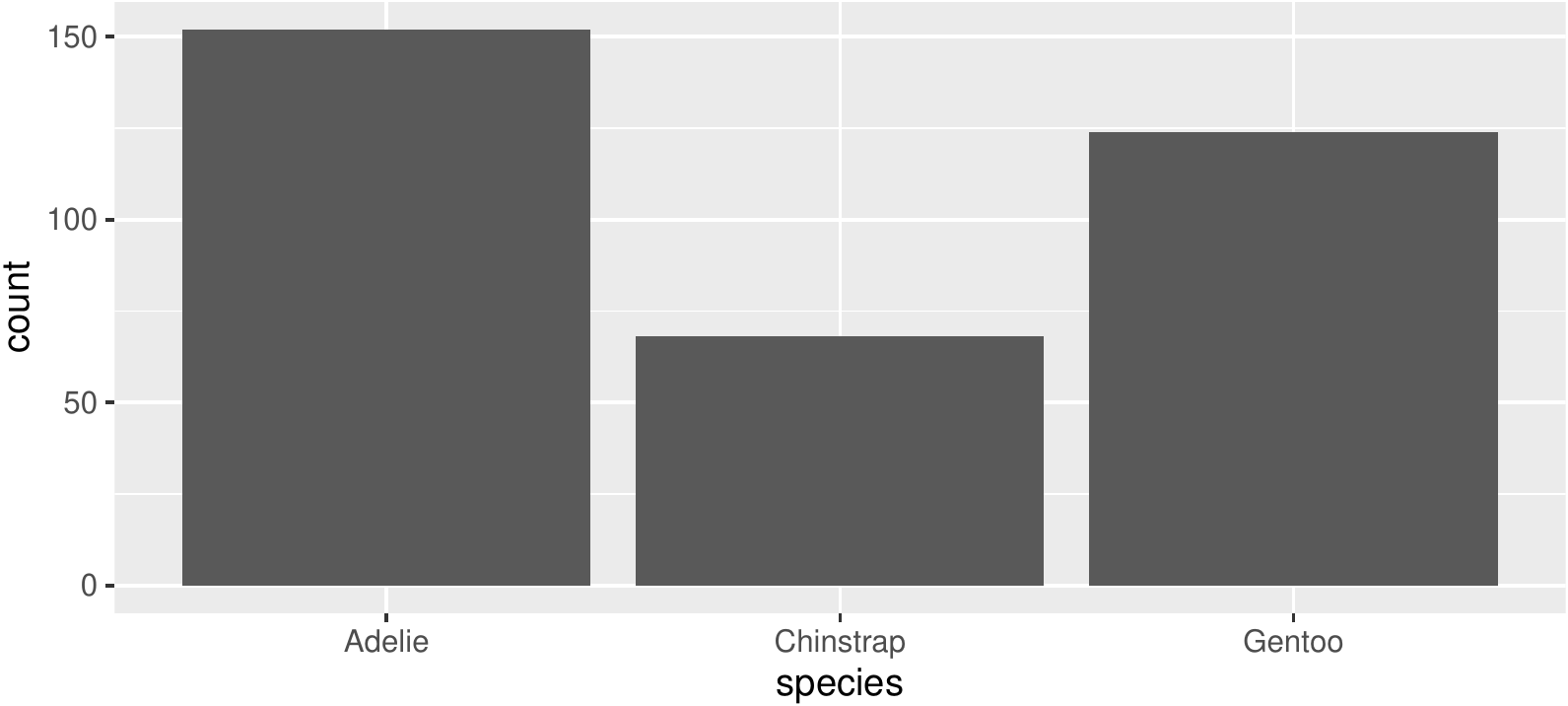} \caption{Sample histogram for automatic data verbalization.}\label{fig:VI-example}
\end{figure}

There are two ways to create alt text: (1) auto-generated alt text; and (2) manual alt text. Currently, \texttt{BrailleR::VI()} \citep{R-BrailleR} function has a capability to auto-generate alt text for basic R graphics and ggplot objects (see Figure \ref{fig:VI-example}).
Some of the supported graphic types include: \texttt{graphics::hist()}, \texttt{graphics::boxplot()}, \texttt{ggplot2::geom\_bar()}, \texttt{ggplot2::geom\_histogram()}, and \texttt{ggplot2::geom\_boxplot()}.

\begin{Shaded}
\begin{Highlighting}[]
\FunctionTok{library}\NormalTok{(ggplot2)}
\FunctionTok{library}\NormalTok{(BrailleR)}
\FunctionTok{library}\NormalTok{(palmerpenguins)}

\CommentTok{\# Create graph}
\NormalTok{g\_species }\OtherTok{\textless{}{-}} \FunctionTok{ggplot}\NormalTok{(}\AttributeTok{data =}\NormalTok{ penguins, }\AttributeTok{mapping =} \FunctionTok{aes}\NormalTok{(}\AttributeTok{x =}\NormalTok{ species)) }\SpecialCharTok{+}
  \FunctionTok{geom\_bar}\NormalTok{()}

\NormalTok{g\_species}
\end{Highlighting}
\end{Shaded}

\begin{verbatim}
## This is an untitled chart with no subtitle or caption.
## It has x-axis 'species' with labels Adelie, Chinstrap and Gentoo.
## It has y-axis 'count' with labels 0, 50, 100 and 150.
## The chart is a bar chart with 3 vertical bars.
## Bar 1 is centered horizontally at Adelie, and spans vertically from 0 to 152.
## Bar 2 is centered horizontally at Chinstrap, and spans vertically from 0 to 68.
## Bar 3 is centered horizontally at Gentoo, and spans vertically from 0 to 124.
\end{verbatim}

\begin{Shaded}
\begin{Highlighting}[]
\CommentTok{\# Obtain auto verbalization}
\CommentTok{\# (Added feature in BrailleR \textgreater{}= 0.32.1): \textasciigrave{}BrailleR::VI()\textasciigrave{} function is automatically called against supported graphics if BrailleR package is currently loaded in R session}
\CommentTok{\# BrailleR::VI(g\_species)}
\end{Highlighting}
\end{Shaded}

Even though the automated alternate text has some use with its limitations, it does not often directly convey the message a visualization is displaying.
Thus, it is important to write alternate text manually and teach students how to write one.
We rely on Cesal's ``simple formula for writing alt text for data visualization'' \citeyearpar{cesal}, which is as follows: ``alt= \emph{Chart type} of \emph{type of data} where \emph{reason for including chart}.''
She also recommends the inclusion of a link to data somewhere within the text.
Similarly, for a meaningful alternate text, \citet{canelon} suggest that the description should convey the meaning of the data, variables on the axes, scale, and type of plot should be included.

Given these recommendations, we provide the alternate text we wrote for Figure \ref{fig:penguins-basic}. ``Sample scatterplot showing the relationship between flipper length in mm on the x-axis and bill length in mm on the y-axis. Flipper lengths vary from about 170 to 230, and bill lengths vary from about 35 to 60. Overall there is a moderate positive relationship. Each data point is colored differently for three species as Adelie, Chinstrap, and Gentoo. Adelie has a short flipper length and short bill length in comparison. Chinstrap has low flipper length but high bill length. Gentoo has a high flipper length and high bill length. Data points are in red, green, and blue for Adelie, Chinstrap, and Gentoo penguins, respectively.''
Note that we specify colors as they would be relevant as readers switch from Figure \ref{fig:penguins-basic} to Figure \ref{fig:penguins-okabe}.

Once alt text is prepared either programmatically or manually, it can be embedded in static and dynamic plot objects in R, such as \texttt{ggplot2::labs(...,\ alt\ =\ "{[}alt\ text{]}")},
\texttt{shiny::renderPlot(...,\ alt\ =\ "{[}alt\ tex{]}")} \citep{R-shiny}. We also ask our students to provide alt text within their R Markdown chunks by using \texttt{fig.alt\ =\ "{[}alt\ text{]}"} \texttt{knitr} chunk option \citep{knitr2014}, and within Pandoc-flavor/Common Markdown with \texttt{!{[}alt\ text\ here.{]}(path/to/file)} syntax \citep{macfarlanePandoc2022}.

\hypertarget{data-sonification-using-sound}{%
\subsection{Data sonification using sound}\label{data-sonification-using-sound}}

\begin{figure}
\includegraphics{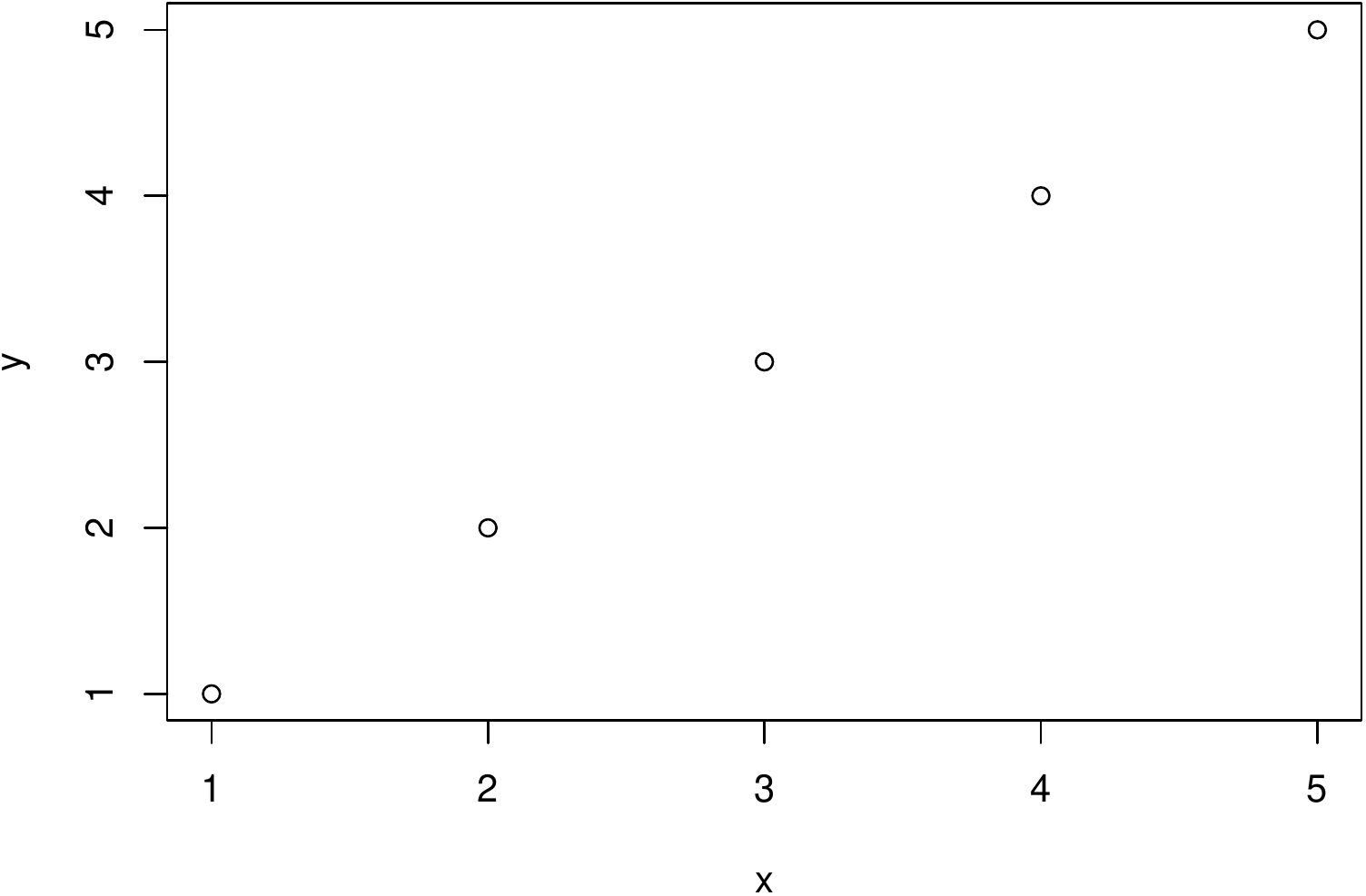} \caption{Visualizing the relationship of two numerical variables with scatter plot.}\label{fig:plot-xy}
\end{figure}

Our definition of sonification refers to a data representation method using stereo sound and various audible patterns. For example, Figure \ref{fig:plot-xy} can be made audible by using \texttt{sonify} package \citep{R-sonify} like below. The output is audio which can also be accessed as an audio file at \url{https://jooyoungseo.github.io/teaching-accessibility-manuscript/audio/sonify-example.wav}.

\begin{Shaded}
\begin{Highlighting}[]
\CommentTok{\# Representing the relationship between the two numerical variables via stereo sound}
\FunctionTok{library}\NormalTok{(}\FunctionTok{sonify}\NormalTok{())}

\CommentTok{\# Prepare variables}
\NormalTok{x }\OtherTok{\textless{}{-}} \DecValTok{1}\SpecialCharTok{:}\DecValTok{5}
\NormalTok{y }\OtherTok{\textless{}{-}} \DecValTok{1}\SpecialCharTok{:}\DecValTok{5}

\FunctionTok{sonify}\NormalTok{(x, y)}
\end{Highlighting}
\end{Shaded}

In this example, a scatter plot is represented by sound. Position on the X axis is communicated using the left-to-right stereo effect; the pitch of the sound indicates the position on the Y axis.

Data sonification is very useful when representing linear patterns; however, there is much room for improvement in terms of complex data sonification as of this writing. Thus, we suggest using sonification for simple linear regression models.

\hypertarget{data-tactualization-using-swell-form-machine}{%
\subsection{Data tactualization using Swell Form machine}\label{data-tactualization-using-swell-form-machine}}

\begin{figure}

{\centering \includegraphics[height=0.5\textheight]{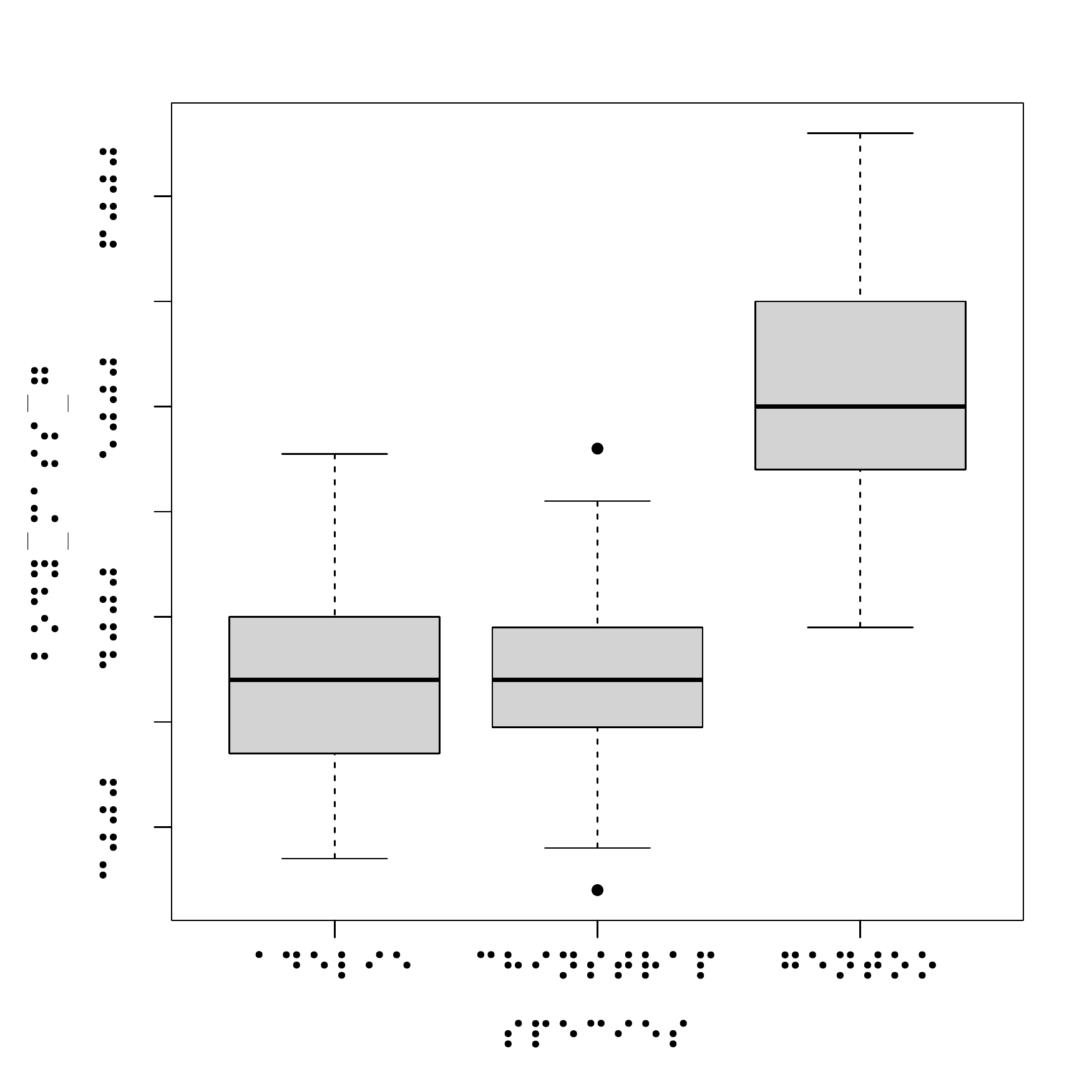} 

}

\caption{Ready-to-print tactile graph in braille.}\label{fig:tactile-example}
\end{figure}

Data tactulization refers to making data visualization in a tactile form so that it can be touchable by fingers.
Although this requires some expensive embossing hardware (e.g., braille embosser or Swell Form Heating Machine), this may be one of the most effective and accessible data representations for those who are blind or visually impaired.
The \texttt{tactileR} \citep{R-tactileR} package in R can generate a ready-to-emboss pdf file from R graphics and ggplot objects.
Students and instructors can produce their own tactile graphics by using the following functions: \texttt{tactileR::brl\_begin()}; {[}their graphic object{]}; \texttt{tactileR::brl\_end()} (see Figure \ref{fig:tactile-example}).
For those who do not have access to embossing hardware, the learning process on tactualization can be supported with a video that shows the process of printing a tactile graph available at \url{https://www.youtube.com/watch?v=ClI555l4Z1M&ab_channel=JooYoungSeo}.

\begin{Shaded}
\begin{Highlighting}[]
\FunctionTok{library}\NormalTok{(palmerpenguins)}
\FunctionTok{library}\NormalTok{(tactileR)}

\FunctionTok{brl\_begin}\NormalTok{(}\StringTok{"figure/tactile{-}boxplot.pdf"}\NormalTok{)}
\FunctionTok{with}\NormalTok{(penguins, }\FunctionTok{boxplot}\NormalTok{(body\_mass\_g }\SpecialCharTok{\textasciitilde{}}\NormalTok{ species))}
\FunctionTok{brl\_end}\NormalTok{()}
\end{Highlighting}
\end{Shaded}

\hypertarget{delivery-of-content}{%
\section{Delivery of Content}\label{delivery-of-content}}

We believe that a natural connection between data science and accessibility can be achieved in the data science classroom while presenting data visualization and then introducing the other aforementioned forms of data representation.
While delivering the content, instructors should try to provide as many opportunities as possible for students to experience using these tools themselves.
For instance, each student should use a screen reader at least once regardless of their visual ability status and hear the alternate text that they have written.

Ample opportunities in learning accessibility should also include assessments.
For instance, instructors incorporating accessibility should modify assessment instructions and rubrics to include accessibility.
This will prevent students from just hearing about accessibility in the lecture and then forgetting about it.
Ideally, the assessments should not treat accessibility as a learning objective at a single point in the academic term.
For instance, if accessibility is covered in Week 3 of the term, students can (and should) still be expected to write alternate texts for visualizations in Week 10.

In this manuscript, we have focused on domain-specific (i.e., data science) knowledge of accessibility using a particular language (R).
Depending on the courses taught, instructors may want to extend the accessibility content to non-data contexts.
In addition, they may adopt the content in their courses with other languages such as Python and SAS or even in their language-agnostic courses.
Overall, if taught well, a student learning accessibility content, for instance, should be able to include alternate text in their PowerPoint presentation in their Anthropology class.
In other words, accessibility knowledge should extend beyond data science and beyond any tool.

Although we have mainly focused on tools and different representations of data to have students develop accessible content, a more crucial accessibility feature is reproducible workflows.
Scholars often focus on reproducibility and replication of scientific findings.
We believe reproducibility can often serve as an accessibility feature in a data science context.
A reproducible open-source workflow allows users who are blind or have a visual impairment to examine data in their preferred form of data representation.
Thus, in addition to teaching different representations of data, instructors may also consider teaching reproducibility.

\hypertarget{closing-remarks}{%
\section{Closing Remarks}\label{closing-remarks}}

We believe that an accessible future in data science requires educators to teach accessibility as part of the curriculum deliberately.
We also acknowledge that what we have presented here is only a small portion of the intersection of visual accessibility and data science.
The second co-author was supported through a Teach Access grant to develop curricular materials on teaching accessibility as part of the data science curriculum.
Educators interested in teaching accessibility can follow the organization's work at \url{https://teachaccess.org/}.
The reproducible R source code of this manuscript can be found on a GitHub repository at \url{https://github.com/jooyoungseo/teaching-accessibility-manuscript}.

\bibliographystyle{agsm}
\bibliography{bib/references.bib,bib/packages.bib}

\end{document}